\documentclass[twocolumn,showpacs,preprintnumbers,amsmath,amssymb,prl]{revtex4}
\usepackage{color}
\usepackage{bm, graphicx, amsmath}
\begin{document}

\title{Extracting information from a qubit by multiple observers: Toward a theory of sequential state discrimination}
\author{Janos Bergou$^{1}$, Edgar Feldman$^{2}$, and Mark Hillery$^{1}$}
\affiliation{$^{1}$Department of Physics, Hunter College of the City University of New York, 695 Park Avenue, New York, NY 10065 USA}
\affiliation{$^{2}$Department of Mathematics, Graduate Center of the City University of New York, 365
Fifth Avenue, New York, NY 10016 USA}

\begin{abstract}
We discuss sequential unambiguous state-discrimination measurements performed on the same qubit.  Alice prepares a qubit in one of two possible states.  The qubit is first sent to Bob, who measures it, and then on to Charlie, who also measures it.  The object in both cases is to determine which state Alice sent.  In an unambiguous state discrimination measurement, we never make a mistake, i.e.\ misidentify the state, but the measurement may fail, in which case we gain no information about which state was sent.  We find that there is a nonzero probability for both Bob and Charlie to identify the state, and we maximize this probability.  The probability that Charlie's measurement succeeds depends on how much information about the state Alice sent is left in the qubit after Bob's measurement, and this information can be quantified by the overlap between the two possible states in which Bob's measurement leaves the qubit.  This paper is a first step toward developing a theory of nondestructive sequential quantum measurements, which could be useful in quantum communication schemes. 
\end{abstract}

\pacs{03.67.-a}

\maketitle

When an observer performs a standard projective quantum measurement on a system, the state of the system after the measurement, the so-called post-measurement state, $|\phi\rangle$, is an eigenstate of the operator that was measured. The measurement is, thus, destructive, and it is generally assumed that any information about the initial state, $|\psi\rangle$, of the system is lost in this process. If, immediately after the first measurement, a second observer performs another measurement on the system the results are describable in terms of $|\phi\rangle$, the post-measurement state of the first observer and not in terms of the initial state $|\psi\rangle$. Therefore it is generally assumed that consecutive measurements on the same quantum system do not yield additional information on the initial preparation, because every consecutive observation prepares the system in a new state.  

The purpose of this paper is to show that this commonly accepted view of standard quantum measurements can be very significantly refined. We show that it is possible to perform consecutive observations on the same system by multiple observers in such a way that each observer in the chain obtains information about the initial state. In fact, and this is the most surprising of our findings, we show that it is possible that each observer obtains {\it full} information about the state in which the system was prepared initially.

We illustrate these ideas for the case in which there are two observers in the observation chain, and each of them performs an unambiguous state discrimination measurement. We emphasize that this is for illustrative purposes only, the same ideas work for more than two consecutive observers and other measurement scenarios. 

In its simplest form unambiguous state discrimination (UD) is the following measurement task. Alice prepares a qubit in one of two known states, $|\psi_{1}\rangle$ or $|\psi_{2}\rangle$, and sends it to Bob. His task is to determine what the state of the qubit is, with no error permitted \cite{ivanovic,dieks,peres,bergou}.  If $|\psi_{1}\rangle$ and $|\psi_{2}\rangle$ are not orthogonal, Bob cannot succeed all the time; the price to pay for no error is that the measurement that distinguishes the states will sometimes fail.  That is, the measurement has three possible outcomes, $1$, corresponding to $|\psi_{1}\rangle$, $2$, corresponding to $|\psi_{2}\rangle$, and $0$, corresponding to failure or inconclusive outcome. The measurement is optimal if the probability of failure is a minimum and is given by $|\langle\psi_{1}|\psi_{2} \rangle |$ in the case that the states are equally probable.  UD is employed in, e.g., quantum key distribution, quantum secret sharing and quantum algorithms \cite{bennett,mimih,bhh}.  It has also been implemented experimentally using the polarization states of photons \cite{huttner,clarke}.

Here we address the question whether more than one user can identify the initial state of the same qubit.  In this scenario Alice prepares a qubit in either $|\psi_{1}\rangle$ or $|\psi_{2}\rangle$ and sends it to Bob.  Bob performs an unambiguous discrimination measurement on the qubit, and sends it on to Charlie, who also performs an unambiguous discrimination measurement on the qubit.   We want both Bob and Charlie to have a nonzero chance of identifying the state so that the probability of both of them succeeding is a maximum.  The rules of the game are that any pre-measurement conspiracy is allowed among all parties but no classical communication can take place between Bob and Charlie after Bob performs his measurement, a scenario typical in secure quantum communication strategies. So, in particular, Charlie never knows whether Bob's measurement succeeded or failed.  The key to making this procedure work is that the state discrimination Bob performs cannot be optimal, otherwise he would have extracted all of the quantum information carried by the qubit, and there would be none left for Charlie to measure.  

Thus, the question of how much information about a state is left after it has been measured is more subtle that is commonly assumed, especially if the measurement is a generalized one, which is described by a POVM (Positive Operator Valued Measure).  However, some information is left even in the case of projective measurements. Rap\v{c}an \emph{et al.} \cite{rapcan} examined how a second observer could ``scavenge'' information about a quantum state that has previously been measured by a first observer.  In their scenario, the second observer has no information about the measurement made by the first, and yet he is still able to gain information about the initial state of the system.  In our scenario, Charlie knows exactly what type of measurement Bob will perform.  Without this condition  Charlie would not be able to perform unambiguous discrimination.

To begin we assume that Alice prepares qubits in $|\psi_{1}\rangle$ or $|\psi_{2}\rangle$ with equal probability. Without loss of generality, the overlap of the two possible states, $s=\langle\psi_{1}|\psi_{2}\rangle$ is taken to be real ($0\leq s \leq 1$) and we choose the phase of $|\psi_{1}^{\perp}\rangle$, the vector orthogonal to $|\psi_{1}\rangle$, so that
\begin{eqnarray}
|\psi_{2} \rangle & = & s|\psi_{1}\rangle + \sqrt{1-s^{2}}|\psi_{1}^{\perp}\rangle \nonumber \\
|\psi_{2}^{\perp}\rangle & = & \sqrt{1-s^{2}}|\psi_{1}\rangle - s |\psi_{1}^{\perp}\rangle .
\end{eqnarray}

Both Bob's and Charlie's measurements are described by POVM's \cite{nielsen}.  Each POVM has three elements, one, $\Pi_{1}$, corresponding to the detection of $|\psi_{1}\rangle$, the second, $\Pi_{2}$, corresponding to the detection of $|\psi_{2}\rangle$, and the third, $\Pi_{0}$, corresponding to the failure of the measurement.  Each element is a positive operator on the two-dimensional qubit Hilbert space, and their sum is the identity operator.  If one is measuring a qubit in the state $|\psi_{i}\rangle$, the probability of obtaining the outcome $j$ is $\langle \psi_{i}|\Pi_{j} |\psi_{i}\rangle$.

The requirement that errors are not allowed mandates that the POVM elements describing Bob's measurement are of the form $\Pi_{1}^{B} = c_{1}|\psi_{2}^{\perp}\rangle\langle\psi_{2}^{\perp}|$ and  $\Pi_{2}^{B} = c_{2}|\psi_{1}^{\perp}\rangle\langle\psi_{1}^{\perp}|$ for the conclusive outcomes and 
\begin{equation} 
\Pi_{0}^{B} = I-\Pi_{1}^{B}-\Pi_{2}^{B}
\label{Pizero1}     
\end{equation}
for the inconclusive one, since the three elements add to the identity. Here $c_{1}$ and $c_{2}$ are positive constants yet to be determined, subject to the constraint $\Pi_{0} \geq 0$. $\Pi_{1}$ and $\Pi_{2}$ are positive by construction.  

The probability that Bob unambiguously detects $|\psi_{i}\rangle$ if it is sent is given by $p_{i}=\langle\psi_{i}|\Pi_{i}^{B}|\psi_{i}\rangle$, for $i=1,2$ and the probability that the measurement fails if $|\psi_{i}\rangle$ is sent is given by $q_{i}=\langle\psi_{i}|\Pi_{0}^{B}|\psi_{i}\rangle$.  Note that the probability that $|\psi_{j}\rangle$ is detected if $|\psi_{i}\rangle$ is sent is zero for $i \neq j$, so $p_{i}+q_{i}=1$.  These relations allow us to express $c_{i}$ in terms of the more physical success and failure probabilities,
\begin{equation}
c_{i} = \frac{p_{i}}{1 - s^{2}} = \frac{1 - q _{i}}{1 - s^{2}} \ .
\label{civsqi}
\end{equation}

We will have to know the states after Bob's measurement, since they will be the input states for Charlie's measurement. They can be expressed in terms of the so-called detection operators $A_{j}$ that are related to the corresponding POVM elements by $\Pi_{j}^{B}=A_{j}^{\dagger}A_{j}$ for $j=0,1,2$.   If $|\psi_{i}\rangle$ is the state  before the measurement, then if we obtain the result $i$ for the measurement ($i=1,2$ success), the post-measurement state (success state) $|\phi_{i}\rangle$ is given by
\begin{equation}
|\phi_{i}\rangle = \frac{A_{i}|\psi_{i}\rangle}{\| A_{i}\psi_{i}\|} ,
\end{equation}
and if we obtain the result $0$ for the measurement, the post-measurement state (failure state) $|\chi_{i}\rangle$ is given by
\begin{equation}
|\chi_{i}\rangle = \frac{A_{0}|\psi_{i}\rangle}{\| A_{0}\psi_{i}\|} ,
\end{equation}
The operators $A_{j}$ can be chosen in the form $A_{j}=U_{j}(\Pi_{j}^{B})^{1/2}$, where $U_{j}$ can be any unitary 
operator.  Thus, we have quite a bit of freedom in choosing these operators and, consequently, Bob's post-measurement states.  In our case they can be expressed as $A_{1}= \sqrt{c_{1}}|\phi_{1}\rangle\langle\psi_{2}^{\perp}|$ and $A_{2}= \sqrt{c_{2}}|\phi_{2}\rangle\langle\psi_{1}^{\perp}|$.

We can now see what happens after Bob's measurement.  If Alice sent $|\psi_{i}\rangle$, then Bob will send Charlie the state $|\phi_{i}\rangle$ with probability $p_{i}$ or the state $|\chi_{i}\rangle$ with probability $q_{i}$. However, we know that for unambiguous discrimination to be possible, the states to be discriminated must be linearly independent \cite{chefles}, and since we are in a two-dimensional space, Charlie can only discriminate between two possible pure states.  This mandates the choice $|\phi_{i}\rangle = |\chi_{i}\rangle$ which, in turn, implies  
\begin{equation}
A_{0}  =  \sqrt{a_{1}} |\phi_{1}\rangle\langle\psi_{2}^{\perp}| +  \sqrt{a_{2}} |\phi_{2}\rangle\langle\psi_{1}^{\perp}| \ ,
\label{Azero}
\end{equation}
where $a_{1}$ and $a_{2}$ are constants to be determined.
Therefore, if Alice sent $|\psi_{1}\rangle$, Charlie will receive $|\phi_{1}\rangle$, whether Bob's measurement succeeded or not, and if Alice sent $|\psi_{2}\rangle$, Charlie will receive $|\phi_{2}\rangle$, again whether Bob's measurement succeeded or not.  Charlie's task, then, is to optimally discriminate between $|\phi_{1}\rangle$ and $|\phi_{2}\rangle$. Further, since $\langle \psi_{i}|A_{0}^{\dagger} A_{0}|\psi_{i}\rangle = q_{i}$, we have that 
\begin{equation}
a_{i} = q_{i}/(1 - s^{2}) \ .
\label{ai}
\end{equation}

We now have two different expressions for $\Pi_{0}$, Eq. \eqref{Pizero1} and $A_{0}^{\dagger}A_{0}$ from \eqref{Azero}, so we still have to check their compatibility.  In the $\{ |\psi_{1}\rangle , |\psi_{1}^{\perp}\rangle \}$ basis the operator $\Pi_{0}^{B}$, Eq. \eqref{Pizero1}, takes the form
\begin{equation}
\Pi_{0}^{B} = \left( \begin{array}{cc} 1-c_{1} + c_{1}s^{2} & c_{1}s\sqrt{1-s^{2}} \\  c_{1}s\sqrt{1-s^{2}} & 1 - c_{1}s^{2} - c_{2} \end{array}\right) .
\end{equation}
It is easy to obtain the eigenvalues and corresponding eigenvectors explicitly. For our purposes, however, the conditions of non-negativity of $\Pi_{0}$, $\mathrm{Tr}(\Pi_{0}) = 2 - c_{1} - c_{2} \geq 0$ and $\det{\Pi_{0}} = 1 - c_{1} - c_{2} + c_{1}c_{2}(1 - s^{2}) \geq 0$, are more useful. The second is the stronger of the two conditions. When it is satisfied the first one is always met. Using \eqref{civsqi}, the condition on the failure probabilities takes the form,
\begin{equation}
1 \geq q_{1}  q_{2} \geq s^{2} \ .
\label{cond1}
\end{equation}
If we now calculate $\Pi_{0} = A_{0}^{\dagger} A_{0}$ from \eqref{Azero} with $a_{i}$ from  \eqref{ai}, we find that the two expressions agree if 
\begin{equation}
q_{1}  q_{2} = \frac{s^{2}}{t^{2}} \ 
\label{cond2}
\end{equation}
where we introduced $\langle\phi_{1}|\phi_{2}\rangle \equiv t$, which we can assume is real and positive. The condition \eqref{cond2} is clearly compatible with \eqref{cond1} provided $t = \langle\phi_{1}|\phi_{2}\rangle \geq s = \langle\psi_{1}|\psi_{2}\rangle$. 

\begin{figure}[ht,floatfix]
      \centering
      \includegraphics[height=5 cm]{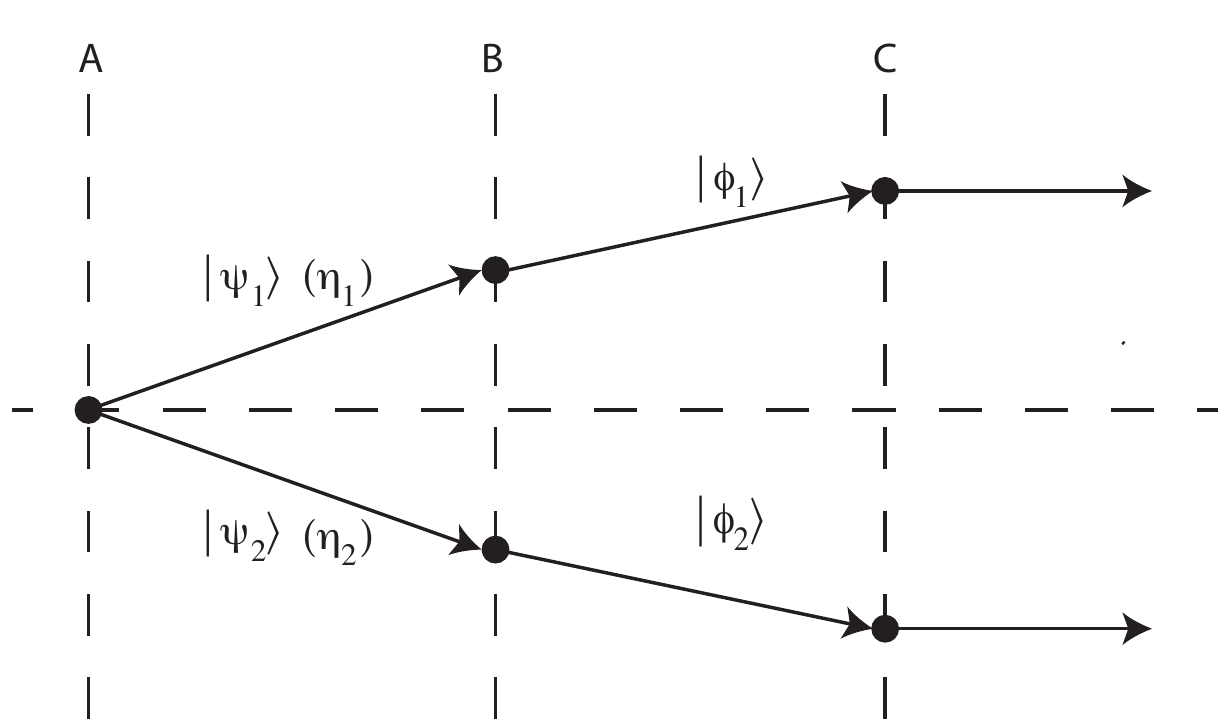}
      \caption{Measurement tree for the sequential measurement. Alice prepares a qubit either in the state $|\psi_{1}\rangle$, which happens with probability $\eta_{1}$ or in the state $|\psi_{2}\rangle$, which happens with probability $\eta_{2}$, such that $\eta_{1} + \eta_{2} = 1$. For simplicity, we assume $\eta_{1} = \eta_{2} = 1/2$. She then hands the qubit to Bob who performs an unambiguous discrimination measurement on it. If he received the qubit in state $|\psi_{1}\rangle$ his post-measurement state will be $|\phi_{1}\rangle$ and if he received the qubit in state $|\psi_{2}\rangle$ his post-measurement state will be $|\phi_{2}\rangle$. The overlap, $t$, of the post-measurement states is increased relative to the overlap, $s$, of the initial states, so $s < t <1$. Bob then sends Charlie the qubit which is now in one of the post-measurement states. Charlie then performs an optimal unambiguous discrimination measurement on the qubit and extracts the remaining information, increasing the overlap of all post-meaurement states to $1$.}
      \label{fig1}
      \end{figure}

The emerging picture is now the following. Bob extracts some information about the two possible inputs, $|\psi_{1}\rangle$ and $|\psi_{2}\rangle$. By doing so he produces states with a greater overlap, $t > s$. Charlie's task, then, is to optimally discriminate between $|\phi_{1}\rangle$ and $|\phi_{2}\rangle$. Since an optimized measurement extracts all of the remaining information, Charlie's post-measurement states can carry no further information about the initial preparation so for all inputs and outcomes they are collapsed to the same common state. The failure probabilities for Bob's measurement must satisfy the constraint given by Eq. \eqref{cond2}. Charlie's failure probabilities must satisfy an entirely similar constraint that we can most easily obtain by replacing $s$ with $t$ and $t$ with $1$ in \eqref{cond2}, since we notice that for his measurement $t$ is the overlap of the input states and the overlap of the post-meassurement states is $1$. The two constraints are given together as [upper index B (C): Bob (Charlie)]
\begin{equation}
q_{1}^{B}  q_{2}^{B} = \frac{s^{2}}{t^{2}} \ , \ \ \ \ \ \ \  q_{1}^{C}  q_{2}^{C} = t^{2}.
\label{cond3}
\end{equation}
The corresponding measurement tree is shown in Fig. \ref{fig1}.

Let us now examine the probability of both measurements succeeding.  Clearly, for the upper branch of the measurement tree in Fig. \ref{fig1} the joint probability of success is $P_{1} = p_{1}^{B}p_{1}^{C} = (1 - q_{1}^{B})(1 - q_{1}^{C})$ and for the lower branch $P_{2} = p_{2}^{B}p_{2}^{C} = (1 - q_{2}^{B})(1 - q_{2}^{C})$, so the average joint success probability is
\begin{equation}
P_{S} = \frac{1}{2}[(1 - q_{1}^{B})(1 - q_{1}^{C}) + (1 - q_{2}^{B})(1 - q_{2}^{C})] .
\label{PS}
\end{equation}
since each branch has a prior probability of $1/2$. 

This is the quantity we want to optimize under the two constraints given in \eqref{cond3}. We will also impose the conditions that the failure probabilities for both states be the same, i.e.\ $q_{1}^{B}=q_{2}^{B}$ and $q_{1}^{C}=q_{2}^{C}$.  Pang et al.\ have shown that for a range of $s$ there are measurements that violate these conditions and give a slightly higher average success probability than those that obey these conditions \cite{pang}. For these measurements, however, the failure probability of one of the states for both Bob and Charlie is $1$, meaning that only one of the two states can be successfully detected.  This renders them impractical for communication purposes, where one needs to be able to detect two alternatives.  In this paper we shall only consider measurements for which the failure probabilities for the two states are the same.  The optimization is now straightforward and can be done by, e.g., using the method of Lagrange multipliers, with the result $q_{1}^{B}=q_{2}^{B}=q_{1}^{C}=q_{2}^{C}=\sqrt{s}$ and $t=\sqrt{s}$. 

Using the optimal values in \eqref{PS}, we finally obtain
\begin{equation}
P_{S}^{(opt)} = (1-\sqrt{s})^{2} \ .
\label{PSopt}
\end{equation}
This equation constitutes the central result of our paper. It clearly shows that there is a finite probability that both of the consecutive observers succeed in extracting the full information about the states.  We also note that the probability of at least one of Bob's or Charlie's measurements succeeding is just $1-s$, which is just the probability of a single optimal unambiguous discrimination measurement of $|\psi_{1}\rangle$ and $|\psi_{2}\rangle$ succeeding. 

So far we have made use of the POVM formalism to describe the unambiguous discrimination measurements.  Another approach is to use the Neumark formalism, in which the system to be measured is coupled to a second system, and projective measurements are performed on the second system.  This type of analysis makes it easier to see what is required for an experimental implementation, and it is discussed in the supplementary material.

\begin{figure}[ht,floatfix]
      \centering
      \includegraphics[height=5 cm]{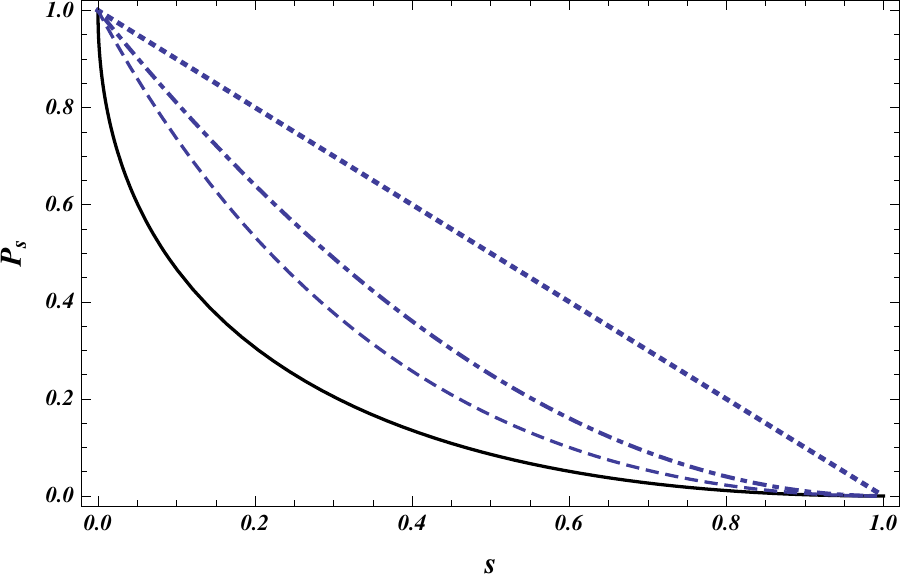}
      \caption{Joint success probability $P_{s}$ vs. $s$ for the four strategies discussed in the paper. Solid line: $P_{S}^{opt}$ vs. $s$, Eq. \eqref{PSopt}. Dotted line: $P_{S}^{(1)}$ vs. $s$, Eq. \eqref{PS1}. Dot dashed line:  $P_{S}^{(2)}$ vs. $s$, \eqref{PS2}. Dashed line: $P_{S}^{(3)}$ vs. $s$, Eq. \eqref{PS3}.}
      \label{fig2}
      \end{figure}

We now want to compare the sequential unambiguous strategy to some strategies that do allow Bob and Charlie to communicate classically.  The strategies to be discussed, like the one discussed above, will not produce any errors.  The strategies are the following.
\begin{enumerate}
\item Bob performs an optimal unambiguous discrimination measurement on the qubit he receives from Alice.  If he succeeds he tells Charlie the results, while if he fails he informs Charlie that his measurement failed, and that is the end of the procedure.  The probability of both of them succeeding is
\begin{equation}
P_{S}^{(1)} = 1- s \ ,
\label{PS1}
\end{equation}
\item Bob performs an optimal unambiguous discrimination measurement on the qubit he receives from Alice.  If he succeeds he sends a qubit in the state he found to Charlie, while if he fails he informs Charlie that his measurement failed, and that is the end of the procedure.  The probability of both of them succeeding is
\begin{equation}
P_{S}^{(2)} = (1- s)^2 \ ,
\label{PS2}
\end{equation}
\item Bob probabilistically clones the qubit he receives from Alice \cite{duan}.  If he succeeds he keeps one clone and sends the other to Charlie, and both apply optimal unambiguous discrimination to their qubits.  If the cloning fails he informs Charlie, and that is the end of the procedure.  The probability that both succeed is
\begin{equation}
P_{s}^{(3)} = (1- s)^{2}/(1+ s) \ ,
\label{PS3}
\end{equation}
\end{enumerate}
The performance of all of the strategies  is compared in Fig. \ref{fig2}.  Finally, let us mention that the probability that at lease one of the parties succeeds is $1-s$, and this is the same for all four strategies.

Therefore, if we only consider the probability of one or both of the parties identifying the state, none of the strategies that allow Bob and Charlie to communicate classically does better than the strategy that does not allow them to communicate.  However, the strategies that allow communication all do better when we consider the probability of both parties identifying the state.  Note that the three protocols enumerated above use more than one qubit, while the sequential unambiguous discrimination protocol uses only one.

The sequential scheme we propose can be generalized in several directions. One obvious generalization is for general prior probabilities. Another one could be the extension to more than two consecutive observers. Instead of just Bob and Charlie one could have B$_{1}$, B$_{2}, \ldots$, B$_{n}$ and there will be a finite probability that each one successfully identifies the initial state of the qubit. The optimal joint probability of success for the case of equal failure probabilities is given by
\begin{equation}
P_{S}^{(opt,n)} = (1-s^{1/n})^{n} \ ,
\label{PSoptn}
\end{equation}
which is a straightforward generalization of \eqref{PSopt}. Finally, the theory of sequential measurements is not at all restricted to POVMs and can be extended to other measurement scenarios including, in particular, standard projective measurements. These and other generalizations are left, however, for a separate publication \cite{bfh}.

In summary, the scheme we have proposed, successive unambiguous discrimination measurements on the same qubit, could be useful in quantum communication schemes.  For example, the B92 quantum cryptography protocol is based on communication using nonorthogonal states \cite{bennett}, and the sequential discrimination scheme could be combined with it to distribute a key to more than one party.    This is discussed further in the supplementary material.

\begin{acknowledgments}
\emph{Acknowledgments}. This research was partially supported by the Institute for Quantum Optics and Quantum Information, University of Vienna (JB) and the National Science Foundation under grant PHY-0903660.
\end{acknowledgments}

\section*{Supplementary material}
\subsection*{Neumark approach}
In the main body of the paper, we described the measurement performed by Bob using the POVM formalism.  Here we would like to describe it via the Neumark formalism.  This was first done in \cite{pang}.  Any POVM measurement can be achieved by coupling the system to be measured to an ancilla, evolving the joint system, and performing projective measurements on the ancilla.  This is just the Neumark description of the POVM measurement, and it is useful in seeing what needs to be done in order to design an experimental implementation of the measurement.

In our case, because our measurement has three outcomes, the qubit we are measuring needs to be coupled to a qutrit.  We shall denote the space of the qubit by $\mathcal{H}_{a}$, which is spanned by the orthonormal basis $\{ |0\rangle , |1\rangle \}$, and the space of the qutrit by $\mathcal{H}_{b}$, which is spanned by the orthonormal basis $\{ |0\rangle , |1\rangle , |2\rangle \}$.  We will choose the initial state of the ancilla to be $|0\rangle_{b}$, and the projective measurement will be performed in the orthonormal basis of the qutrit.  The result $|0\rangle_{b}$ will correspond to failure, the result $|1\rangle_{b}$ will correspond to $|\psi_{1}\rangle$, and the result $|2\rangle_{b}$ will correspond to $|\psi_{2}\rangle$.  The unitary transformation coupling the qubit and qutrit, $U$, must then have the action
\begin{eqnarray}
\label{U}
U|\psi_{1}\rangle_{a}|0\rangle_{b} & = & |\phi_{1}\rangle_{a} (\sqrt{1-q_{1}}|1\rangle_{b} + \sqrt{q_{1}} |0\rangle_{b}) \nonumber \\
U|\psi_{2}\rangle_{a}|0\rangle_{b} & = & |\phi_{2}\rangle_{a} (\sqrt{1-q_{2}}|1\rangle_{b} + \sqrt{q_{2}} |0\rangle_{b}) .
\end{eqnarray}
This form guarantees that Bob's measurement will never produce an error, and that Charlie will get the same state no matter whether Bob's measurement succeeds or fails

We can make this more specific if we choose particular states for $|\psi_{j}\rangle$ and $|\phi_{j}\rangle$, $j=1,2$
\begin{eqnarray}
|\psi_{1}\rangle_{a} & = & \cos\theta |0\rangle_{a} + \sin\theta |1\rangle_{a} \nonumber \\
|\psi_{2}\rangle_{a} & = & \cos\theta |0\rangle_{a} - \sin\theta |1\rangle_{a} ,
\end{eqnarray}
and 
\begin{eqnarray}
|\phi_{1}\rangle_{a} & = & \cos\theta^{\prime} |0\rangle_{a} + \sin\theta^{\prime} |1\rangle_{a} \nonumber \\
|\phi_{2}\rangle_{a} & = & \cos\theta^{\prime} |0\rangle_{a} - \sin\theta^{\prime} |1\rangle_{a} .
\end{eqnarray} 
This implies that $s$, the overlap between $|\psi_{1}\rangle$ and $|\psi_{2}\rangle$, is given by $s=\cos (2\theta )$, and that $t$, the overlap between $|\phi_{1}\rangle$ and $|\phi_{2}\rangle$, is given by $t= \cos (2\theta^{\prime})$.  We will consider the case in which the probability of both measurements succeeding is a maximum, which implies that $q_{1}=q_{2} = t = \sqrt{s}$.  Substituting the specific forms for the states into Eqs.\ (\ref{U}), we can find the action of $U$ on the states $|0\rangle_{a}|0\rangle_{b}$ and $|1\rangle_{a}|0\rangle_{b}$.  Defining the vectors 
\begin{eqnarray}
|v_{1}\rangle_{b} & = & \frac{1}{(1+\sqrt{s})^{1/2} } \left[ \sqrt{2}s^{1/4}|0\rangle_{b} \right. \nonumber \\
& & \left. +\left(\frac{1-\sqrt{s}}{2}\right)^{1/2}(|1\rangle_{b}+|2\rangle_{b})\right] \nonumber \\
|v_{2}\rangle_{b} & = & \frac{1}{\sqrt{2}}(|1\rangle_{b} - |2\rangle_{b} ) ,
\end{eqnarray}
we find that 
\begin{eqnarray}
U|0\rangle_{a} |0\rangle_{b} & = & \frac{1}{[2(1+s)]^{1/2}} \left[ (1+\sqrt{s}) |0\rangle_{a} |v_{1}\rangle_{b} \right. \nonumber \\
& & \left. + (1-\sqrt{s}) |1\rangle_{a}|v_{2}\rangle_{b} \right] \nonumber \\
U|1\rangle_{a}|0\rangle_{b} & = & \frac{1}{\sqrt{2}} (|0\rangle_{a}|v_{2}\rangle_{b} + |1\rangle_{a}|v_{1}\rangle_{b} ) .
\end{eqnarray}
These equations, then, specify the action of a unitary operator that could be used in the implementation of the unambiguous discrimination measurement discussed in the main body of the paper.

\subsection*{Alternate strategies}
The discussion of alternate strategies was rather abbreviated in the main text, so more details are provided here.  The first strategy was for Bob to perform an optimal unambiguous discrimination measurement on the qubit he receives from Alice.  If he succeeds he tells Charlie the results, while if he fails he informs Charlie that his measurement failed, and that is the end of the procedure.  In this case, if Bob fails, they both fail and if Bob succeeds they both succeed.  The probability of Bob succeeding is 
\begin{equation}
P_{S}^{(1)} = 1- s \ ,
\label{PS1}
\end{equation}
which is also the probability of both of them succeeding. This strategy, allowing full classical communication after Bob's measurement, is expected to have the highest probability of both parties succeeding.  
  
The second strategy was for Bob to perform an optimal unambiguous discrimination measurement on the qubit he receives from Alice.  If he succeeds he sends a qubit in the state he found to Charlie, while if he fails he informs Charlie that his measurement failed, and that is the end of the procedure.   In this case, if Bob fails, they both fail.  The probability of at least one of them succeeding is the same as that of Bob succeeding, $1-s$, while the probability of both of them succeeding is 
\begin{equation}
P_{S}^{(2)} = (1- s)^2 \ ,
\label{PS2}
\end{equation}
This strategy has the same probability of at least one of the parties succeeding, but a smaller probability of joint success than the previous one because there is less classical communication.  

The third strategy was for Bob to probabilistically clone the qubit he receives from Alice.  A probabilistic cloner can produce perfect clones of an input qubit if the state of that qubit is chosen from a limited set.   In addition, it does not always succeed, but we do know when it has succeeded.  If the input qubit can only be in either $|\psi_{1}\rangle$ or $|\psi_{2}\rangle$, then the probability of successfully producing two clones at the output is $1/(1+s)$.  The cloning strategy consists of Bob cloning the qubit he gets from Alice, and if the cloning is successful, he keeps one qubit and sends the other to Charlie.  If the cloning fails, he tells Charlie, and that is the end of the procedure.  Both Bob and Charlie perform optimal unambiguous discrimination measurements on their qubits.  The probability that at least one of the parties learns which state Alice sent, which is the probability that the cloning succeeds and that at least one of the measurements succeeds is $1-s$. This is the same as in the two previous strategies.  The probability that both parties determine which state Alice sent, which is the probability that the cloning succeeds and both measurements succeed, is 
\begin{equation}
P_{s}^{(3)} = (1- s)^{2}/(1+ s) \ ,
\label{PS3}
\end{equation}
which is greater than $(1-\sqrt{s})^{2}$, but less than $(1-s)^{2}$ since there is some classical communication but less than in the previous cases.

\subsection*{B92 protocol}
In the B92 protocol, Alice sends a qubit in one of two nonorthogonal states, $|\psi_{1}\rangle$ or $|\psi_{2}\rangle$, to Bob, with $|\psi_{1}\rangle$ corresponding to the bit value $0$ and $|\psi_{2}\rangle$ corresponding to the bit value $1$.  Bob measures the states using the optimal unambiguous discrimination measurement for these two states.  Bob tells Alice, for each of the qubits sent, whether the measurement succeeded or failed, and the probability that it succeeded is $1-s$.  For those cases in which it succeeded, Alice and Bob share a bit.  The cases is which Bob's measurement failed are discarded.  An eavesdropper, Eve, is faced with the problem of discriminating two nonorthogonal states, and, since this cannot be done perfectly, she will invariably introduce errors.  Alice and Bob compare a subset of the bits on which Bob's measurement succeeded, and if there are no errors, there was no eavesdropping.

Suppose now Alice wants to share key bits with Bob and Charlie.  One possibility is that she can send each of them a qubit in the same state, either both in $|\psi_{1}\rangle$ or both in $|\psi_{2}\rangle$.  This procedure requires two qubits, and the probability of both Bob's and Charlie's measurements succeeding is $(1-s)^2$.  A second possibility is that Alice sends one qubit to Bob, who performs the unambiguous discrimination measurement proposed in this paper and then sends the same qubit on to Charlie.  In that case, only one qubit is required, but the probability of both measurements succeeding, in the optimal case, is $(1-\sqrt{s})^2$.  The first strategy has a higher success probability, while the second is cheaper in terms of qubits.

A second difference between the two strategies is in their susceptibility to certain eavesdropping attacks.  Suppose Eve employs unambiguous discrimination in her eavesdropping attempts.  In the two-qubit scenario, Eve can measure each of the two qubits that Alice sends, so her probability of learning which state was sent is $1-s^2$.  In the one-qubit scenario, Eve's best choice is to capture the qubit between Alice and Bob, and apply a optimal unambiguous discrimination measurement.  Her probability of success is $1-s$.  If she succeeds, she knows the state, but if she fails and makes a guess as to which state to send on to Bob, there is no point in her attacking the link between Bob and Charlie, because it provides her with no additional information.  So, under this simple kind of attack, each round of the single-qubit strategy is less susceptible to eavesdropping than each round of the two-qubit strategy.  Clearly a much more detailed analysis would have to be done in order to determine under what conditions each strategy is best, but this is something that will have to be reserved for future work.

\end{document}